\documentclass[sigconf]{acmart}

\usepackage{booktabs}
\usepackage{graphicx}
\usepackage{amsmath}
\usepackage{hyperref}
\usepackage{algorithm}
\usepackage{algpseudocode}
\usepackage{soul}
\usepackage{tikz}
\usetikzlibrary{positioning, arrows.meta}
\usepackage[utf8]{inputenc}

\begin{document}

\title[Tiled MatMul Accelerator for Transformer Self-Attention]%
      {Design and Implementation of an FPGA-Based Tiled Matrix Multiplication Accelerator for Transformer Self-Attention on the Xilinx KV260 SoM}

\author{Richie Li}
\authornote{Legal name: Zhaoqin Li}
\affiliation{%
  \institution{University of California, Irvine}
  \city{Irvine}
  \country{United States}
}
\email{zhaoqil3@uci.edu}

\author{Sicheng Chen}
\affiliation{%
  \institution{University of California, Irvine}
  \city{Irvine}
  \country{United States}
}
\email{sichenc5@uci.edu}

\begin{abstract}
Transformer-based large language models (LLMs) rely heavily on intensive matrix multiplications for attention and feed-forward layers, with the Q, K, and V linear projections in the Multi-Head Self-Attention (MHA) module constituting a decisive performance bottleneck. In this work, we introduce a highly optimized tiled matrix multiplication accelerator on a resource-constrained Xilinx KV260 FPGA that not only addresses this challenge but sets a new standard for efficiency and performance. Our design exploits persistent on-chip storage, a robust two-level tiling strategy for maximal data reuse, and a systolic-like unrolled compute engine that together deliver unparalleled speed and energy efficiency. Integrated with DistilBERT for Q, K, and V projections, our accelerator achieves an unequivocal \textbf{7×} speedup over ARM CPU implementations (PyTorch) and an extraordinary \textbf{200×} improvement over naive NumPy, reaching a throughput of up to 3.1~GFLOPs for matrix multiplications on (64,768) × (768,3072) matrices while operating at a conservative 100\,MHz. These results decisively demonstrate the transformative potential of FPGA-based acceleration for critical Transformer operations, paving the way for scalable and energy-efficient deep learning inference on edge devices.
\end{abstract}



\maketitle

\section{Introduction}
\begin{figure*}
    \centering
    \includegraphics[width=1\linewidth]{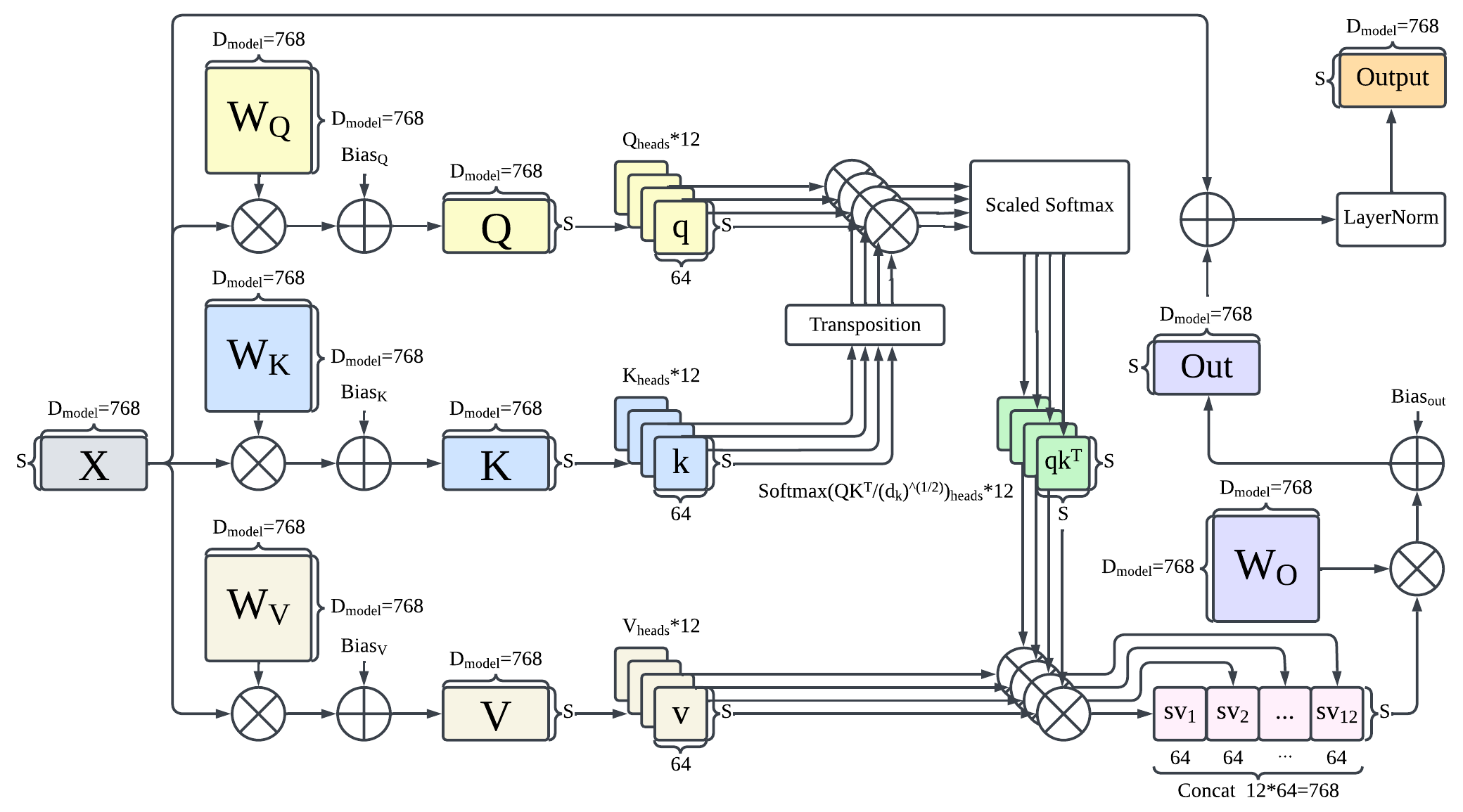}
    \Description{Diagram showing the multi-head self-attention pipeline in a Transformer model.}
    \caption{MHA Pipeline in transformer}
    \label{fig:mha_pipeline}
\end{figure*}

Large Language Models (LLMs) like Transformers rely on intensive matrix multiplications for their self-attention and feed-forward blocks. As shown in the Figure ~\ref{fig:mha_pipeline}, computing the Q, K, V, and output linear projections during Multi-Head Self-Attention in each Transformer layer entails multiplying large matrices (input activation $\times$ Wq/Wk/Wv/Wo). DistilBERT~\cite{distilbert}, a compact Transformer variant~\cite{vaswani2017attention}, still performs billions of multiply-accumulate operations per inference. These operations are memory-bandwidth heavy on general processors, motivating specialized hardware acceleration. FPGAs offer custom parallelism and energy efficiency for matrix math, but deploying LLM acceleration on edge FPGAs presents challenges due to limited resources and power.

Motivated by the observation that the Q, K, and V projections form a significant computational bottleneck, we strategically concentrated our efforts on accelerating these key components. This targeted approach allowed us to fully optimize our design for the resource constraints of the Xilinx Kria KV260 Vision AI Kit (housing a Zynq UltraScale+ FPGA), while still delivering substantial performance improvements. Our approach exploits data reuse and parallelism to speed up matrix multiplication operations within the tight DSP and BRAM budget of an edge FPGA. We adopt a block/tiled matrix multiply approach: one input matrix (``A'') remains in on-chip memory, while the other (``B'') is processed in smaller column blocks to maximize reuse. Unlike prior LLM accelerators that apply model-wide optimizations (e.g., pruning, specialized sparse formats), our work directly accelerates dense GEMM (General Matrix Multiply) cores within the Q, K, V projection operations. To encourage reproducibility and further research, we provide our FPGA design, software interface, and benchmarking scripts in an open-source repository at \textbf{\href{https://github.com/Richielee630/TMMA}{GitHub}}.

\section{Contributions}
\label{sec:contributions}

In this work, we present a specialized FPGA-based accelerator for key matrix multiplication operations in Transformer self-attention. While elements of our contributions are woven throughout the \emph{Introduction} and \emph{Discussion} sections, here we explicitly summarize the main novel aspects and potential benefits for future research:

\begin{itemize}
    \item \textbf{Targeted Focus on Self-Attention Bottleneck:} 
    We pinpoint the Q, K, and V projections within Multi-Head Self-Attention as critical performance bottlenecks. Unlike many existing FPGA accelerators that aim to cover a broad range of model operations (often at the cost of higher resource demands), our design is tightly focused on these fundamental matrix multiplications, making it more suitable for edge FPGAs with constrained resources.

    \item \textbf{Tiled and Data-Centric Design:}
    By employing two-level tiling and on-chip data persistence, we minimize off-chip memory transfers and fully leverage data reuse. Our systolic-like unrolled compute engine—sized specifically to fit KV260 resources—enables high throughput (up to 3.12 GFLOPs) while staying within strict DSP and BRAM budgets. This data-centric approach is amenable to analytic frameworks such as MAESTRO~\cite{MAESTRO} and other DNN dataflow studies~\cite{DNNDataflows}.

    \item \textbf{High-Level Synthesis (HLS) and Easy Integration:}
    Implementing the accelerator in C++ HLS facilitates rapid design space exploration and more accessible customization. The accelerator is integrated via AXI4 and PYNQ overlays, offering a user-friendly interface that can be extended to other matrix-multiply-based workloads with minimal effort.

    \item \textbf{Quantized DistilBERT Integration:}
    We replace standard PyTorch Q, K, and V linear layers with our \texttt{FPGAQuantizedLinear} module and validate end-to-end functionality on a quantized DistilBERT. This demonstrates near-lossless accuracy while providing up to a 7$\times$ speedup over PyTorch CPU baselines for these layers. Our open-source release includes scripts for quantization, FPGA invocation, and reproducible benchmarking, potentially lowering the barrier for researchers to adopt FPGA acceleration in transformer applications.

    \item \textbf{Roadmap for Future Scaling:}
    Although we focus on moderate matrix sizes (e.g., 64$\times$768, 768$\times$3072) representative of DistilBERT, our tiled architecture offers a blueprint for scaling to larger transformer variants. The clear separation of on-chip buffering, tiling logic, and computation modules can be extended to handle more extensive hidden dimensions and different attention heads with minimal redesign. 

\end{itemize}

By addressing a critical compute bottleneck within Transformers and providing an open-source HLS-based implementation, we aim to advance both the practical adoption of FPGAs in edge-deployed large language models and future studies on optimized dataflows for deep learning inference.

\section{Related Work}
\subsection*{FPGA Accelerators for Transformers}
Recent research has produced FPGA-based LLM accelerators such as FlightLLM~\cite{FlightLLM} and SSR~\cite{SSR}. \textbf{FlightLLM} (FPGA’24) maps entire LLM inference flows onto FPGAs, leveraging sparsity and mixed-precision; implemented on a high-end Alveo U280, it achieves \textbf{6× higher energy efficiency} than an NVIDIA V100 GPU~\cite{FlightLLM} by using sparse DSP chains and always-on-chip decoding. \textbf{SSR} (FPGA’24) explores launching multiple accelerators in parallel versus sequentially to balance latency and throughput. On a Versal ACAP VCK190, SSR attains up to \textbf{2.5× throughput} versus an NVIDIA A10G GPU, with \textbf{8.5× energy efficiency}~\cite{SSR}. \textbf{FAMOUS} focuses on the Transformer’s attention mechanism: a flexible multi-head attention core on Alveo U55C that sustains 328~GOPS and runs \textbf{2.6× faster than an NVIDIA V100 GPU} and \textbf{1.3× faster} than prior FPGA designs. Earlier, Qi \emph{et al.} (GLSVLSI’21) combined model compression (pruning) with FPGA optimization to fit Transformers onto FPGAs. These works, however, target large FPGAs or datacenter contexts and often integrate advanced techniques (sparsity, multi-engine concurrency) less applicable to small-edge devices.

In addition to large-scale accelerators, recent research has advanced data-centric analytical frameworks to understand data reuse and mapping strategies in DNN accelerators. For instance, MAESTRO~\cite{MAESTRO} and the approach presented in “Understanding Reuse, Performance, and Hardware Cost of DNN Dataflows: A Data-Centric Approach”~\cite{DNNDataflows} provide comprehensive models for quantifying data movement, reuse, and the associated hardware costs. Although these frameworks target a broader range of DNN workloads, they share foundational principles with our work—particularly the emphasis on maximizing both temporal and spatial data reuse—which informs our targeted accelerator design for Transformer self-attention on edge FPGAs.

\subsection*{Our Approach vs Prior Art}
Unlike FlightLLM or SSR, which assume abundant resources (HBM memory, $>9000$ DSPs) and apply complex optimizations, our design emphasizes \textbf{dense GEMM acceleration on a resource-constrained FPGA}. We focus on core matrix-multiply throughput via tiling and on-chip buffering, rather than algorithmic sparsification. In contrast to FAMOUS’s multi-head pipeline on big FPGAs, we implement a single GEMM core optimized for reuse and run it at 100~MHz on the KV260 (which has far fewer BRAM/DSP). Recent work on complete Transformer accelerators, such as the Hardware Accelerator for Multi-Head Attention and Position-Wise Feed-Forward in the Transformer~\cite{MultiHeadAccelerator}, also highlights the importance of efficient dataflow management. However, by concentrating on an efficient memory hierarchy and loop-level optimizations specifically for the Q, K, and V projections, our design is uniquely tailored for edge deployment scenarios.

\section{System Design}
\begin{figure} [H]
    \centering
    \includegraphics[width=1\linewidth]{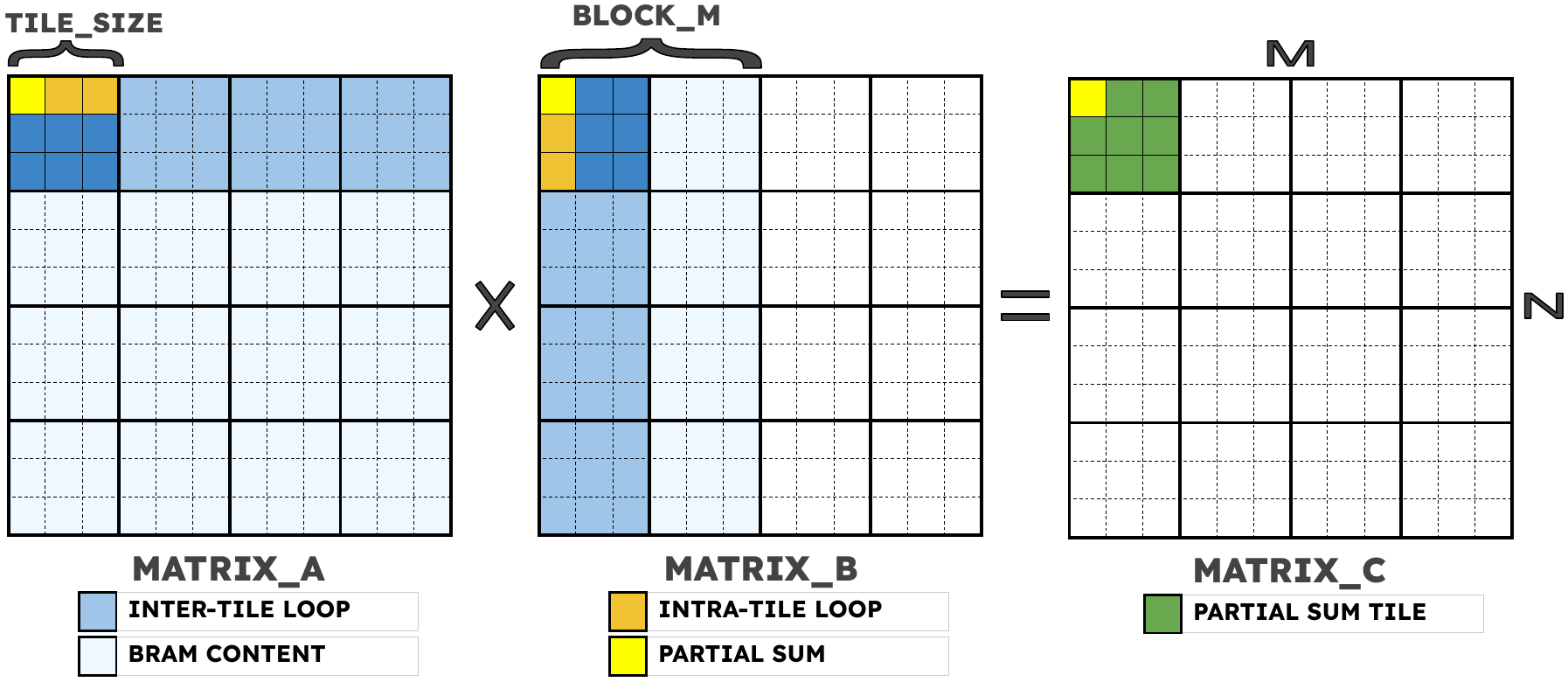}
    \Description{Diagram showing our tiled matrix multiplication approach.}
    \caption{Overview of tiled matrix multiplication approach.\protect}
    \label{fig:tiling_strat}
\end{figure}

The accelerator is based on a \textbf{tiled matrix multiplication architecture} tailored for multiplying matrices of size 64$\times$768 (A) with 768$\times$3072 (B), matching the dimensions used in DistilBERT. By decomposing large matrices into smaller tiles that can be loaded into fast on-chip memory (BRAM), the design improves data locality, reduces off-chip DRAM accesses, and enables extensive parallelism. Figure~\ref{fig:tiling_strat} illustrates an overview of the tiling logic.

\subsection*{High-Level Dataflow Mapping Strategies}
Our design leverages multi-level mapping strategies to optimize both computation and data movement:

\paragraph{I. Temporal Mapping:}
\begin{itemize}
    \item \textbf{Tiling:} We employ a two-level tiling approach (Block Tiling with $M_b=256$ and Inner Tiling with $T=32$) to decompose large matrices into smaller tiles. This allows matrix A to be loaded once into on-chip BRAM, significantly reducing DRAM accesses.
    \item \textbf{Pipelining:} The accumulation loop over the K dimension is pipelined (II=1), ensuring continuous processing and high throughput.
    \item \textbf{Data Persistence:} Persistent storage of matrix A enhances data reuse across multiple computations.
\end{itemize}

\paragraph{II. Spatial Mapping:}
\begin{itemize}
    \item \textbf{Loop Unrolling:} Fully unrolling the innermost loops creates a 32$\times$32 array of multipliers-adders for parallel computation.
    \item \textbf{Memory Partitioning:} Partitioning local tile arrays into registers enables simultaneous data access, reducing memory access bottlenecks.
\end{itemize}

\paragraph{III. Inter/Intra-tile Mapping:}
\begin{itemize}
    \item \textbf{Inter-tile:} Matrix B is partitioned into 256-column blocks, with each block assigned to a different processing column-tile with corresponding row-tile from Matrix A for parallel execution.
    \item \textbf{Intra-tile:} Within each tile, pipelined and fully unrolled loops enable efficient parallel processing of tiles.
\end{itemize}


Together, these mapping strategies optimize the balance between computation and data movement, enabling high throughput on the resource-constrained Xilinx KV260.

Recent data-centric approaches, such as those in MAESTRO~\cite{MAESTRO} and the work on DNN dataflows~\cite{DNNDataflows}, provide an analytical perspective on how temporal and spatial mapping can be exploited to maximize data reuse. Inspired by these frameworks, our design explicitly separates temporal mapping—through techniques like tiling, pipelining, and data persistence—from spatial mapping—through loop unrolling and memory partitioning. This dual strategy minimizes off-chip accesses and maximizes parallelism, thereby achieving high efficiency on an edge FPGA platform.

\subsection{HLS Implementation and Parallel Computation}
At the implementation level, the high-level mapping strategies are realized using Vivado HLS directives:

\begin{itemize}
    \item \textbf{Unified Tiling and Memory Hierarchy:}  
    Matrix A is loaded once into persistent BRAM, while matrix B is streamed in blocks. Local buffers (register-level tiling) store 32$\times$32 tiles of A and B for rapid access.
    
    \item \textbf{Parallel Computation:}  
    The innermost loops are fully unrolled, creating a 32$\times$32 array of multiplier-adder units that operate concurrently. The accumulation loop over the K dimension is pipelined with an initiation interval (II) of 1, achieving up to 1024 int8$\times$int8 MAC operations per clock cycle.
    
    \item \textbf{Practical HLS Optimization:}  
    Loop pipelining, unrolling, and array partitioning work in tandem to implement the temporal and spatial mapping strategies in hardware, thereby maximizing parallelism while respecting the FPGA’s resource constraints.
\end{itemize}

\subsection{AXI Interface and Integration}
\begin{figure*}
    \centering
    \includegraphics[width=1\linewidth]{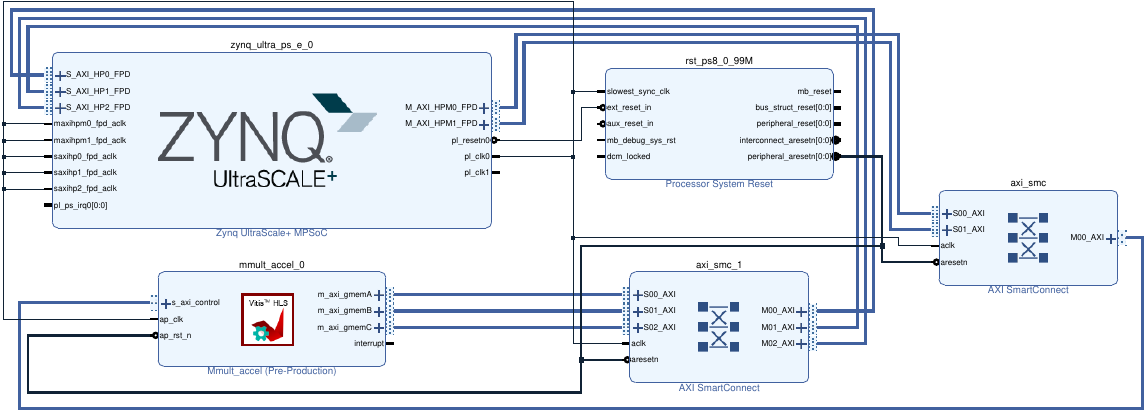}
    \Description{Diagram showing the Vivado Block Design connections beween PS and PL}
    \caption{Vivado Block Design}
    \label{fig:vivado_bd}
\end{figure*}

The accelerator is implemented as an IP core and integrated into the system via AXI SmartConnect interfaces, enabling communication with the Zynq UltraScale+ MPSoC. As shown in Figure~\ref{fig:vivado_bd}, for memory and control, three AXI4 master ports (one each for input A, input B, and output C) stream data to/from external DDR, and an AXI4-Lite slave interface allows the host (CPU) to configure dimensions and start the kernel~\cite{HLS_design}. The design supports a special control flag \texttt{update\_A} so that the host can choose to reuse the last loaded A matrix for subsequent calls~\cite{Quant_forward_pass} – useful when processing multiple B batches with the same weights (e.g., iterating over attention heads or sequences). On the KV260, the accelerator IP is integrated into the FPGA fabric and invoked from software via PYNQ drivers.

\section{Implementation}

\begin{algorithm}[H]
\caption{High-Level Tiled Matrix Multiplication Accelerator}
\label{alg:high_level_mmult}
\begin{algorithmic}[1]
\Require Matrices $A$ ($N\times K$), $B$ ($K\times M$), output matrix $C$ ($N\times M$), flag \texttt{update\_A}; constants: \texttt{BLOCK\_M}, \texttt{TILE\_SIZE}
\Ensure $C \gets A \times B$
\If{\texttt{update\_A} is true}
    \State Copy $A$ from DDR into persistent on-chip BRAM
\EndIf
\For{each column block $j\_block$ in $[0, M)$ with step \texttt{BLOCK\_M}}
    \State $current\_block\_M \gets \min(\texttt{BLOCK\_M}, M - j\_block)$
    \State Load block of $B$ into on-chip BRAM
    \For{each tile row $i_0$ in $[0, N)$ with step \texttt{TILE\_SIZE}}
        \For{each tile column $j_0$ in $[0, current\_block\_M)$ with step \texttt{TILE\_SIZE}}
            \State Initialize local output tile $localC \gets 0$
            \For{each tile $k_0$ in $[0, K)$ with step \texttt{TILE\_SIZE}}
                \State Load $localA$ and $localB$ tile from Bram
                \State $localC \gets localC + localA \times localB$
            \EndFor
            \State Write tile $localC$ to $C$ at the appropriate offset
        \EndFor
    \EndFor
\EndFor
\end{algorithmic}
\end{algorithm}

We developed the accelerator in C++ HLS (Xilinx Vitis HLS). The HLS code explicitly declares the top-level loops and applies pragmas for pipelining and unrolling. For example, the innermost multiply-accumulate loops are annotated with \texttt{\#pragma HLS UNROLL} (for the \texttt{ii} and \texttt{jj} loops over the tile) and the $k$-loop with \texttt{\#pragma HLS PIPELINE II=1}. Similarly, we partition the local tile arrays into registers (\texttt{\#pragma HLS ARRAY\_PARTITION complete}), so each element can be accessed in parallel without memory banking conflicts.

\textbf{Tile size selection.} 
\emph{We experimented with several tile sizes (e.g., $T=16$, $T=64$) and settled on $T=32$ based on routing feasibility, timing closure, and performance.} Larger tiles such as $T=64$ improved theoretical parallelism but made place-and-route challenging at 100~MHz. Smaller tiles (e.g., $T=16$) simplified timing but reduced total throughput. Hence, $T=32$ emerged as an optimal balance between resource usage and achievable frequency.

\textbf{Handling partial tiles.} 
\emph{In real-world cases, $N$, $K$, or $M$ may not be perfectly divisible by 32. We use boundary checks within our tiled loops to handle leftover rows and columns, loading and computing smaller sub-tiles. In our experiments, overhead for these partial tiles is minimal (\textasciitilde1--2\% time difference), indicating our design scales well to arbitrary matrix dimensions.}

The accelerator IP was compiled to RTL with Vivado HLS and integrated in Vivado along with AXI SmartConnect for data movement. We ran at a conservative 100~MHz PL clock, easily meeting timing closure. During synthesis, \emph{some int8 multiplications were mapped to LUTs} once available DSP blocks were exhausted, reflecting the high level of parallelism (up to 1024 concurrent MACs).

\begin{figure}[H]
    \centering
    \includegraphics[width=0.5\linewidth]{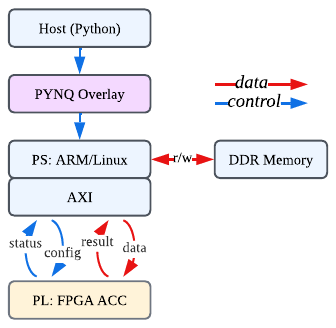}
    \Description{Diagram showing the highlevel overview of the PYNQ Layer Interface}
    \caption{Diagram of the PYQ-based interface on the KV260.}
\label{fig:pynq_overlay_diagram}
\end{figure}

On the software side, we created a PYNQ overlay in Figure ~\ref{fig:pynq_overlay_diagram} to manage buffers and FPGA invocation. The host (Python) configures the accelerator via the PYNQ overlay and the Processing System (PS), as depicted in the diagram of the PYNQ-based interface on the KV260—where the AXI interface provides bidirectional communication (Config/Status, Result) between the PS and the FPGA Accelerator, and DDR memory holds buffers for A, B, and C. Specifically, the host allocates contiguous buffers for A, B, and C (using \texttt{pynq.allocate}), sets up the accelerator registers (pointing to physical addresses of buffers and dimensions N, K, M), and toggles \texttt{AP\_START}~\cite{Quant_forward_pass}. We wrapped this in a Python \texttt{call\_fpga()} function that optionally retains A between calls. This allowed integration with a quantized DistilBERT: we replaced the PyTorch linear layers for Q, K, V in attention with calls to the FPGA (using custom PyTorch extensions to invoke \texttt{call\_fpga()}), feeding int8 quantized weights and inputs. In doing so, we ensured consistent scaling by using symmetric quantization with a fixed scale factor and zero-point for the weights and activations so that the int8 results could be dequantized to match the FP32 baseline. We verified that the FPGA outputs matched CPU computation exactly for small test matrices and remained within quantization error for end-to-end model outputs.

\section{Performance Evaluation}
\subsection{Experimental Setup}
We benchmarked the accelerator on the KV260 board running Ubuntu with PYNQ support. The host's CPU consists of 4$\times$ Arm Cortex-A53 @ 1.5GHz and 2$\times$ Cortex-R5F @ 600MHz, which we used for running baseline software matrix multiplication (NumPy and PyTorch) and orchestrating FPGA execution.

\textbf{Power measurement methodology.} 
\emph{We used the KV260’s onboard power sensors to obtain instantaneous readings every 50--100~ms during execution. Energy consumption was approximated by integrating (summing) the power samples over the total runtime of the matrix multiplication. We then compared this to the CPU baseline, similarly sampling the system power with the FPGA accelerator quiescent.}

Latency was measured using Python’s \texttt{time()} function along with the accelerator’s built-in timers.

\begin{table}[ht]
\centering
\caption{Resource Utilization for Tiled MatMul Accelerator on XCK26 (KV260) at 100\,MHz}
\label{tab:res_util}
\begin{tabular}{lccc}
\toprule
\textbf{Resource} & \textbf{Used} & \textbf{Available} & \textbf{Utilization} \\
\midrule
BRAM   & 126        & 144        & 88\% \\
DSP48E & 1040       & 1248       & 83\% \\
FF     & 102741     & 237600     & 43\% \\
LUT    & 71050      & 118800     & 60\% \\
\bottomrule
\end{tabular}
\end{table}

\subsection{Benchmark Cases}
We evaluated two scenarios:
\begin{enumerate}
    \item \textbf{Standalone GEMM:} Random matrices of varying sizes were used to assess raw computational performance. We conducted tests on two representative cases:
    \begin{itemize}
        \item A DistilBERT attention case of (64, 768) $\times$ (768, 768), corresponding to approximately 37.7 million MACs.
        \item A Feed-Forward Network (FFN) case of (64, 768) $\times$ (768, 3072), corresponding to roughly 150 million MACs.
    \end{itemize}
    \item \textbf{DistilBERT Attention Throughput:} Here, we integrated the FPGA accelerator into a quantized DistilBERT model by replacing the standard Q, K, and V linear layers with our custom \texttt{FPGAQuantizedLinear} module. This module offloads only the critical Q, K, V matrix multiplications from the Multi-Head Self-Attention (MHA) block.
\end{enumerate}

\begin{table}[ht]
\centering
\begin{tabular}{lcc}
\toprule
\textbf{Framework} & \textbf{Latency (s)} & \textbf{Throughput (GFLOPs)} \\
\midrule
NumPy (ARM CPU)   & 20.72  & 0.01 \\
PyTorch (ARM CPU) & 0.67   & 0.45 \\
FPGA (compute)    & 0.09   & 3.12 \\
FPGA (end-to-end) & 0.11   & 2.85 \\
\bottomrule
\end{tabular}
\caption{Performance on a 768$\times$3072 matrix multiplication.}
\label{tab:performance}
\end{table}

For the 768$\times$3072 multiplication, the FPGA achieved a latency of 9.67~ms compared to 67.84~ms on PyTorch (ARM) and 2072.25~ms on NumPy (without optimized BLAS), yielding a \textbf{7.0× speedup} over PyTorch and a \textbf{214× speedup} over NumPy. The effective throughput was approximately 3.12~GFLOPs for the core computation, slightly reducing to 2.85~GFLOPs when including data transfer overhead.

Energy measurements further underscored the benefits of our approach: the KV260 consumed about 3.3~W during FPGA operation compared to a baseline of 3~W in idle mode, while the ARM core used roughly 3.2~W when running PyTorch. This resulted in an energy efficiency of approximately 0.5~J per 768$\times$3072 multiply on the FPGA versus 2~J on the ARM—a roughly 4× improvement.

In the \textbf{DistilBERT Attention} scenario, our focus was on accelerating the Q, K, and V linear projections. The \texttt{FPGAQuantizedLinear} module performs the following:
\begin{itemize}
    \item Quantizes input activations and weights to int8.
    \item Converts these arrays into PYNQ buffers for efficient data transfer.
    \item Offloads the core 2D matrix multiplication to the FPGA accelerator.
    \item Dequantizes the resulting int32 outputs back to floating point and adds bias if necessary.
\end{itemize}
Benchmarking compared the inference times of a CPU-only forward pass against those with FPGA acceleration for these projections. Our experimental results show that while a CPU-only forward pass required approximately 1.14 seconds, offloading the Q, K, and V projections reduced the compute time for matrix multiplications to about 0.43 seconds. Consequently, the overall end-to-end speedup for DistilBERT inference is around 2×. Although the transmission latency introduced by the PYNQ overlay tempers the speedup relative to the standalone GEMM performance, the nearly identical prediction confidence levels (e.g., 99.95\% on CPU versus 99.80\% on FPGA) decisively demonstrate that our approach robustly accelerates a critical component of the attention mechanism without compromising accuracy.

\section{Challenges and Optimizations}

During development, we encountered several bottlenecks and addressed them via optimized HLS transformations:

\begin{itemize}
    \item \textbf{Memory Bandwidth Bottleneck:} Initially, reading A and B from DRAM each time was a limiting factor. We introduced \textbf{persistent BRAM buffers} for A and block BRAM for B to cut down redundant transfers~\cite{HLS_design}. Additionally, we used wide AXI bursts to transfer blocks of B efficiently (256 columns at a time). The \texttt{update\_A} flag allows us to amortize the cost of loading A when it remains constant.

    \item \textbf{HLS Pipeline Timing:} Achieving II=1 in the inner loop was challenging given 1024 operations in parallel. We ensured no loop-carried dependencies by fully unrolling the independent MAC operations. The HLS tool initially issued warnings about DSP utilization; \emph{we mitigated this by allowing some multipliers to be mapped to LUTs once DSPs were fully consumed,} which still met timing requirements.

    \item \textbf{Loop Bounds and Edge Conditions:} We implemented tile loops to cover matrix sizes not multiples of 32 (with boundary checks inside the tile loads and compute). \emph{Testing revealed that partial-tile overhead was low, so no specialized hardware was needed for fractional tiles.}

    \item \textbf{Persistent On-Chip Storage:} Storing the entire A (64$\times$768 int8) in BRAM consumed about 48~KB, which was acceptable. However, storing a full 768$\times$768 B (approximately 589~KB) was impossible, hence the block tiling. We tuned \texttt{BLOCK\_M=256} after evaluating BRAM usage.

    \item \textbf{Quantization:} Converting DistilBERT weights and activations to int8 required careful calibration. We used PyTorch’s static quantization to get int8 weights for the Q, K, V linear layers. Our quantized accelerator produced negligible accuracy loss (<0.5\% deviation in attention outputs).

    \item \textbf{Clock Frequency vs Parallelism Tradeoff:} 
    \emph{We attempted tile sizes of $T=16$ and $T=64$ to explore the design space. $T=16$ reduced concurrency (fewer multipliers in parallel), lowering throughput. $T=64$ improved concurrency but complicated place-and-route, failing timing closure at 100~MHz. Thus, $T=32$ offered a well-balanced solution.}
\end{itemize}

\section{Discussion and Future Work}
The project demonstrates that a relatively small FPGA can accelerate key operations of Transformer models. By deliberately focusing on the acceleration of the Q, K, and V linear projections within the Multi-Head Self-Attention module, we have optimized a critical performance bottleneck under resource-constrained conditions. The substantial improvements observed in standalone GEMM benchmarks indicate that extending this targeted approach to other components could yield even more significant end-to-end gains. Our results echo the insights from broader data-centric frameworks~\cite{MAESTRO, DNNDataflows} that emphasize the benefits of maximizing temporal and spatial data reuse.

\begin{itemize}
    \item \textbf{Focused Acceleration:} Our accelerator currently targets the Q, K, and V projections—a deliberate choice that allowed for deep optimization and efficient use of resources. Future work can build on this foundation by integrating additional components, such as softmax operations and the Feed-Forward Network (FFN) layers.
    \item \textbf{Scaling to Larger Models:} Our design currently targets matrices with input activation sizes up to 64$\times$768. For larger Transformer models (e.g., with hidden sizes of 1024 or 4096), advanced memory management strategies—such as double-buffered streaming and ping-pong buffering—will be essential to hide latency and effectively manage data flow.
    \item \textbf{Reducing Data Transfer Overhead:} To unlock the FPGA’s full potential, future work should further minimize data movement between the CPU and FPGA. Leveraging more persistent on-chip memory for activations and implementing DMA could significantly reduce transfer latency.
    \item \textbf{Improving CPU-FPGA Execution Pipelining:}  
    Maximizing performance requires overlapping CPU and FPGA execution. While the FPGA effectively accelerates the QKV projections, concurrent processing of subsequent stages (e.g., attention softmax or FFN layers) on the CPU is essential for further throughput gains. Techniques such as double buffering can facilitate simultaneous data transfer and computation, thereby reducing idle time. Addressing the unignorable transmission delays associated with the current PYNQ overlay and custom forward pass represents a promising direction for future work.
    \item \textbf{Optimizing FPGA Utilization and Scaling:} Future improvements might include parallelizing the Q, K, and V projections on separate systolic arrays, as well as keeping the FPGA accelerator active across multiple layers rather than resetting registers and flushing memory for each new computation. Such strategies would reduce overhead and further improve efficiency.
\end{itemize}

\section{Conclusion}
We have presented the design and implementation of an FPGA-Based Tiled Matrix Multiplication Accelerator for Transformer Self-Attention on the Xilinx KV260 SoM. Our work demonstrates that through loop tiling, on-chip buffering, and parallel unrolling, even a resource-constrained FPGA can significantly accelerate the matrix multiplication operations that underpin Transformer attention mechanisms. Standalone GEMM benchmarks and the acceleration of Q, K, and V projections achieved a notable 7$\times$ speedup over PyTorch CPU execution and marked improvements in energy efficiency.

While the overall end-to-end DistilBERT performance currently exhibits a 2$\times$ speedup, this result is largely influenced by the transmission delays inherent to the current PYNQ overlay and custom forward pass. These unignorable overheads clearly indicate an opportunity for further optimization. Future work focused on reducing data transfer latencies and integrating additional components—such as softmax and FFN layers—along with enhanced system-level optimizations is expected to yield even greater performance gains. In summary, this project not only confirms the viability of FPGA-based acceleration for key Transformer operations but also establishes a strong foundation for future advancements in efficient, scalable deep learning inference on power-constrained edge devices.

\begin{acks}
The authors would like to thank Professor Sitao Huang from the University of California, Irvine, for his invaluable guidance and insightful discussions throughout this project. Special thanks are also extended to Sicheng Chen for his excellent collaboration.
\end{acks}


\begin{thebibliography}{9}

\bibitem{MultiHeadAccelerator}
Siyuan Lu et al. 2020.
\textit{Hardware Accelerator for Multi-Head Attention and Position-Wise 
Feed-Forward in the Transformer}.
arXiv:2009.08605.
\url{https://arxiv.org/abs/2009.08605}.

\bibitem{MAESTRO}
Hyoukjun Kwon, Prasanth Chatarasi, Vivek Sarkar, and Tushar Krishna. 2020.
\textit{MAESTRO: A Data-Centric Approach to Understand Reuse, Performance, 
and Hardware Cost of DNN Mappings}.
IEEE Micro, 40(3): xx--yy, 2020.
\url{https://doi.org/10.1109/MM.2020.2985963}.

\bibitem{DNNDataflows}
Hyoukjun Kwon, Prasanth Chatarasi, Michael Pellauer, Angshuman Parashar, 
Vivek Sarkar, and Tushar Krishna. 2019.
\textit{Understanding Reuse, Performance, and Hardware Cost of DNN Dataflows: 
A Data-Centric Approach}.
In MICRO-52, Oct. 12--16, 2019, Columbus, OH, USA.
\url{https://doi.org/10.1145/3352460.3358252}.

\bibitem{vaswani2017attention}
Ashish Vaswani, Noam Shazeer, Niki Parmar, et al. 2017.
\textit{Attention Is All You Need}.
In Proc. of Advances in Neural Information Processing Systems (NeurIPS) 30.
\url{https://arxiv.org/abs/1706.03762}

\bibitem{distilbert}
Victor Sanh, Lysandre Debut, Julien Chaumond, and Thomas Wolf. 2019.
\textit{DistilBERT, a distilled version of BERT}.
arXiv:1910.01108.
\url{https://arxiv.org/abs/1910.01108}

\bibitem{FlightLLM}
Shulin Zeng et al. 2024.
\textit{FlightLLM: Efficient Large Language Model Inference 
with a Complete Mapping Flow on FPGAs}.
In Proceedings of FPGA ’24.

\bibitem{SSR}
Jinming Zhuang et al. 2024.
\textit{SSR: Spatial Sequential Hybrid Architecture for Latency-Throughput 
Tradeoff in Transformer Acceleration}.
In Proceedings of FPGA ’24.

\bibitem{HLS_design}
HLS Design Code.
\url{https://github.com/Richielee630/MatMul_SA.git}.

\bibitem{SAMA_benchmark}
SAMA PYNQ Benchmark Results.
Available at:
\url{https://github.com/Richielee630/TMMA/tree/main/pynq}.

\bibitem{Quant_forward_pass}
Quantized DistilBERT Forward Pass.
Available at:
\url{https://github.com/Richielee630/TMMA/tree/main/pynq}.

\end{thebibliography}
\end{document}